\documentclass[twocolumn,conference]{IEEEtran}
\IEEEoverridecommandlockouts
\usepackage{array}
\usepackage{booktabs}
\usepackage{cite}
\usepackage[font=small]{caption}
\usepackage{amsmath,amssymb,amsfonts}
\usepackage{graphicx}
\usepackage{subcaption}

\usepackage{todonotes}
\usepackage{algorithm}
\usepackage{algorithmic}

\usepackage{textcomp}
\usepackage{multirow}

\usepackage {url}
\usepackage [numbers,sort&compress]{natbib}
\usepackage[super]{nth}
\usepackage{paralist}
\usepackage[letterpaper, margin=0.75in]{geometry}
\usepackage{color}

\usepackage{textcomp}
\def\BibTeX{{\rm B\kern-.05em{\sc i\kern-.025em b}\kern-.08em
    T\kern-.1667em\lower.7ex\hbox{E}\kern-.125emX}}

\newcolumntype{L}[1]{>{\raggedright\arraybackslash}m{#1}}
\newcolumntype{C}[1]{>{\centering\arraybackslash}m{#1}}

\begin{document}

\title{A Generic Framework for Task Offloading in mmWave MEC Backhaul Networks}


\author{
\IEEEauthorblockN{Kyoomars Alizadeh Noghani$^{*}$, Hakim Ghazzai$^{\#}$, and Andreas Kassler$^{*}$\\
\IEEEauthorblockA{\small $^{*}$Karlstad University, Karlstad, Sweden\\ Email: \{kyoomars.noghani-alizadeh, andreas.kassler\}@kau.se\\
$^{\#}$Stevens Institute of Technology, Hoboken, NJ, USA\\ Email: hghazzai@stevens.edu}
{\thanks {\vspace{-0.4cm}\hrule
\vspace{0.1cm} \textcopyright~2018 IEEE. Personal use of this material is permitted. Permission from IEEE must be obtained for all other uses, in any current or future media, including reprinting/republishing this material for advertising or promotional purposes, creating new collective works, for resale or redistribution to servers or lists, or reuse of any copyrighted component of this work in other works.}}
}
\vspace{-0.4cm}}

\maketitle

\begin{abstract}
With the emergence of millimeter-Wave (mmWave) communication technology, the capacity of mobile backhaul networks can be significantly increased. On the other hand, Mobile Edge Computing (MEC) provides an appropriate infrastructure to offload latency-sensitive tasks. However, the amount of resources in MEC servers is typically limited. Therefore, it is important to intelligently manage the MEC task offloading by optimizing the backhaul bandwidth and edge server resource allocation in order to decrease the overall latency of the offloaded tasks. This paper investigates the task allocation problem in MEC environment, where the mmWave technology is used in the backhaul network. We formulate a Mixed Integer NonLinear Programming (MINLP) problem with the goal to minimize the total task serving time. Its objective is to determine an optimized network topology, identify which server is used to process a given offloaded task, find the path of each user task, and determine the allocated bandwidth to each task on mmWave backhaul links. Because the problem is difficult to solve, we develop a two-step approach. First, a Mixed Integer Linear Program (MILP) determining the network topology and the routing paths is optimally solved. Then, the fractions of bandwidth allocated to each user task are optimized by solving a quasi-convex problem. Numerical results illustrate the obtained topology and routing paths for selected scenarios and show that optimizing the bandwidth allocation significantly improves the total serving time, particularly for bandwidth-intensive tasks.





\end{abstract}


\section{Introduction}
\label{sec:intro}
Nowadays, a notable amount of mobile applications and services including video streaming apps and social network services are hosted in distributed data centers. Furthermore, an increasing number of mobile users rely on their own devices to carry out the storage and computation of intensive operations. The ability to offload tasks from a mobile device to the cloud helps in overcoming the resource limitation of the mobile device, saving its energy, and extending its battery life~\cite{8016573}. The aforementioned goals would be achieved at the expense of experiencing a high latency if cloud services are not provided in close proximity. In this regard, Mobile Edge Computing (MEC)~\cite{MEC} has emerged as a new paradigm in which Base Stations (BSs) are integrated with computing, storage, and networking capabilities to deploy parts of cloud services closer to mobile subscribers. Moreover, thanks to new communication technologies such as millimeter-Wave (mmWave), BSs become able to exchange a large volume of data at a higher rate which in turn may significantly enhance the performance of the MEC infrastructure. 

However, MEC confronts some challenges such as the limited resources of edge servers and the user task assignment. For instance, a key challenge is to determine the destination of computation offloading, i.e., either the edge or central cloud server. Although resource allocation to user demands has been the topic of numerous research, it has been less investigated in the context of MEC. The offloading decision in single user MEC systems with a single dedicated edge server has been investigated in~\cite{5445167, Kumar2013}. Authors in~\cite{7762913, 6612005, 7307234} studied finite radio-and-computational resource allocation to mobile users in multi-user MEC systems with a single dedicated edge server. MEC server scheduling in multi-user MEC systems has been the topic of research such as~\cite{7841937, 6945323}. Server selection problem in heterogeneous MEC system is the most related research field to our work. Authors in~\cite{7414063} proposed an optimal user scheduling for offloading the tasks when both the edge and central cloud coexist and each of them has a single server. Ge et al.~\cite{Ge:2012:GTR:2333660.2333724} formulated game theoretical solutions to model and minimize the total energy consumption of mobile users and edge servers. Finally, Dinh et al.~\cite{7914660} proposed an optimization offloading framework to minimize both the task execution latency and the mobile energy consumption when the mobile device is able to allocate tasks to multiple small cell access points.

To the best of our knowledge, the problem of latency efficient task allocation in MEC environment with mmWave backhaul network has not been investigated before. In this paper, we consider a MEC system consisting of a group of BSs, a set of heterogeneous edge servers, a remote data center, and a set of bandwidth-intensive user tasks. The ultimate goal is to find the mapping between servers and the user tasks with respect to computing resources and network constraints such that the total task serving time is minimized. In summary, we make the following contributions:

\begin{itemize}
\item We propose a generic formulation that models the task allocation problem for mmWave backhaul networks. The output of this model determines an optimized network topology, identifies which server is used to process a given offloaded task, finds the routing paths of each user task, and determines the fractions of allocated bandwidth on mmWave backhaul links to each task. The optimization problem is formulated as a Mixed Integer NonLinear Program (MINLP).

\item As the MINLP optimization problem is difficult to solve optimally, an alternative two-phase approach is developed. First, by assuming fixed bandwidth allocation policy and conducting a series of linearization steps, the optimization problem is converted into a Mixed Integer Linear Program (MILP) that optimally determines the backhaul mesh topology and the user task routing paths. Then, a quasi-convex problem is optimally solved to determine the fractions of bandwidth to be allocated to each user task over the links forming its path. We consider minimum rate and hop-by-hop transmission at the backhaul network and adopt two different latency metrics for the bandwidth allocation accordingly.
\end{itemize}

A numerical evaluation shows that, thanks to the optimized bandwidth allocation, the total serving time is notably decreased compared to the one obtained with the fixed bandwidth allocation policy used with the MILP. Furthermore, we investigate the impact of MEC infrastructure and path diversity on total task serving time. The result shows a dramatic decrease of the latency when MEC infrastructure exists at the backhaul network especially when BSs are equipped with more interfaces.

The rest of the paper is organized as follows. Section II introduces the system model. The problem formulation is developed in Section III. The proposed approach is discussed in Section IV. Selected numerical results are presented in Section V. Finally, the paper is concluded in Section VI.


%


\section{System Model}
\label{sec:system-model}
We consider a mmWave network consisting of $N$ BSs where each BS is equipped with $I$ antennas. The geographical coordination of BS node $n \in \{1 \dots N\}$ is denoted by ($X_{n}$, $Y_{n}$) and the mmWave link capacity among two BSs $n,m \in \{1 \dots N\}: n \neq m$ is denoted by $R_{nm}$. The values of $R_{nm}$ are computed based on the average statistics of the channel, i.e., the path loss due to propagation and atmospheric conditions~\cite{7523521}. We assume that a mmWave link can be established between all interfaces of BS nodes $n$ and $m$ if the received signal at the BS is higher than a certain threshold. We denote $\delta_{n,m}$ as follows:
\begin{equation}
   \delta_{n,m}=\left\{
   \begin{array}{ll}
   1, & \hbox{if the link between BS $n$ and} \\
         & \hbox{BS $m$ can be established,} \\
   0, & \hbox{otherwise.}
   \end{array}
   \right.
\end{equation}

Let us denote the link between i$^{th}$ interface of BS $n$ and j$^{th}$ interface of BS $m$ by ($n$,$i$,$m$,$j$) where $n \neq m$.

We assume that $P$ BSs, where $1 \leq P \leq N$, have a wired connection to the cloud network. Likewise, we denote the total number of BSs that are not directly connected to the cloud by $O$ where $O < N$. The same value of latency, denoted by $\theta$, is considered for all links connecting the BSs to the cloud and we assume that it is independent of the task size. 

In this paper, we are interested in the user tasks that need to be processed in external servers. In this case, the BS collects the tasks from its associated user(s) and becomes in charge of them. We denote the initial location of a user task $b$, where $b \in\{1,\dots, B\}$ and $B$ is the total number of users tasks, by the binary variable $T_{n,b}$. Hence, $T_{n,b}=1$ if user task $b$ is initially located at BS $n$. Moreover, we denote the size of each task by $L_{b}$ where $b \in \{1 \dots B\}$. We assume that each task is highly integrated and has to be executed as a whole. Additionally, we assume that splitting a task over multiple paths is not possible. 

To process a task, various resources such as CPU, memory, and storage are required. In this paper and for simplicity, we limit our study to the storage capacity. The developed framework can be easily extended to consider other resources. BSs can either process the user task locally subject to its resource constraints or forward the task to another server (edge or central cloud) to be processed. We assume that the BSs are heterogeneous entities with different storage capacities, $C_{n}$. Then, we indicate by the binary variable $\Pi_n$ if a BS is co-located with a server as follows:
\begin{equation}
   \Pi_{n}=\left\{
   \begin{array}{ll}
   1, &\hspace{-3mm}\hbox{if BS $n$ is co-located with a server,} \\
   0, &\hspace{-3mm}\hbox{otherwise.}
   \end{array}
   \right.
\end{equation}

Hence, the storage capacity of a BS $n$ is denoted by $\Pi_n C_n$. Herein we assume that servers (both edge and cloud servers) can process multiple tasks in parallel if they have enough capacity. Unlike BSs that have finite storage capacity, we assume that the cloud has an infinite one.

\begin{figure}[t]\vspace{+3mm}
\begin{centering}
\includegraphics[width=0.45\textwidth]{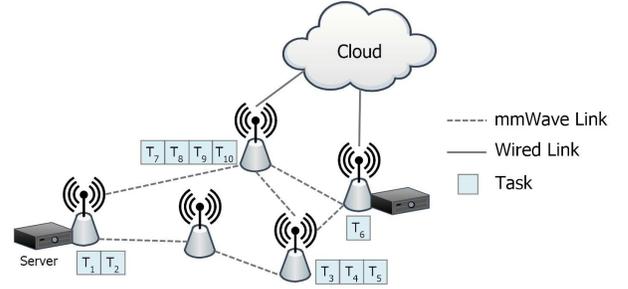}
\par\end{centering}
\caption{\label{fig:system-model}A mmWave backhaul network with local servers, connections to the cloud, and offloaded tasks.}
\vspace{-2mm}
\end{figure}

Fig.~\ref{fig:system-model} illustrates an overview of the system model. A mmWave backhaul network is shown with five BSs where two of the nodes have direct wired connections to the cloud. Additionally, two of the BSs are co-located with edge servers while the other BSs act as relay nodes. Tasks beside each BS show the initial offload points. 

\section{Problem Formulation}
\label{sec:problem-formulation}
This section formulates the task allocation problem minimizing the total task serving latency while considering constraints including the limited capacities of the edge servers and the links. The output of the problem identifies the mapping between a user task and the server which processes the task, finds the routing path of each user task, and determines the fractions of allocated bandwidth to each task in each link. The primary decision variables of the problem are listed in Table~\ref{table:decision-variables}.
\begin{table*}[t]
  \caption{List of decision variables.}
  \label{table:decision-variables}
  \centering
    \begin{tabular}{ | c | c | c | p{10.5cm} |}
    \hline
    Decision Variable & Type & Size & Meaning \\ \hline
    $X_{n,i,m,j}$ & \{0,1\} & $N^2 . I^2$ & equals 1 if the interface $i$ of BS $n$ is connected to interface $j$ of BS $m$ \\ \hline
    $X_{n,i,m,j}^{b}$ & \{0,1\} & $N^2 . I^2 . B$ & equals 1 if the task $b$ is transmitted over the link $(n,i,m,j)$\\ \hline
    $\rho_{n,i,m,j}^{b}$ & (0,1] & $N^2 . I^2 . B$ & defines the allocated portion of the bandwidth to task $b$ on mmWave link $(n,i,m,j)$ \\ \hline
    $Y_{n,b}$ & \{0,1\} & $(N + 1) . B$ & equals 1 if BS $n$ process the task $b$ \\ \hline
    $W_{p,b}$ & \{0,1\} & $P . B$ & equals 1 if the task $b$ is sent to the cloud through BS $p \in \{1, ..., P\}$ \\ \hline
    \end{tabular}
    \vspace{-1mm}
\end{table*}

Note that the size of decision variable $Y$ is $(N + 1)B$ as it contains all $N$ BSs in the mmWave backhaul network plus the remote cloud. 

\subsection{Constraints} \label{Constraints-Section} 
\textbf{Interface Connectivity:} The constraints associated to the decision variables $X(n,i,m,j)$ are given as follows:
\begin{equation}
\label{Interface-Connectivity-1}
\sum_{i=1}^{I}\sum_{j=1}^{I} X_{n,i,m,j} \leq \delta_{n,m},~~~~~~~~~~~~~~~~~~~~~~~~~~~~~~~\forall n, m,
\end{equation}
\begin{equation}
\label{Interface-Connectivity-2}
\sum_{m=1}^{N}\sum_{j=1}^{I} X_{n,i,m,j} + X_{m,j,n,i} \leq 1,~~~~~~~~~~~~~~~~~~~~~~~~~~~\forall n, i.
\end{equation}

Constraint~\eqref{Interface-Connectivity-1} ensures that a mmWave link between node $n$ and $m$ can be established only if $\delta=1$. Constraint~\eqref{Interface-Connectivity-2} guarantees that a given interface of a specific BS is only connected to one interface of another node. Moreover, it ensures that the data is transmitted in one direction.


\textbf{Task Association to Links:} A task can traverse a link if and only if a link exists or is established between the two BSs:
\begin{equation}
\label{Task-Link-1}
\sum_{n=1}^{N}\sum_{i=1}^{I}\sum_{m=1}^{N}\sum_{j=1}^{I} X_{n,i,m,j}^{b} \leq X_{n,i,m,j},~~~~~~~~~~~~~~~~~~~~~~~~~\forall b.
\end{equation}

Furthermore, the following constraint avoids the establishment of redundant links, i.e., when no task is sent over the link:
\begin{equation}
\label{Task-Link-2}
X_{n,i,m,j} \leq \sum_{b=1}^{B} X_{n,i,m,j}^{b},~~~~~~~~~~~~~~~~~~~~~~~~~\forall n, i, m, j.
\end{equation}

\textbf{Flow Conservation:} Constraints~\eqref{Flow-Conservation-1} and~\eqref{Flow-Conservation-2} ensure that a task is forwarded completely by intermediate BSs. Constraint~\eqref{Flow-Conservation-1} indicates that all tasks (the received tasks plus the initially located ones) ought to leave~\textit{or} be processed at that BS. Constraint~\eqref{Flow-Conservation-2} expresses the same limitation for BSs that have connectivity with the cloud.
\begin{subequations}
\label{Flow-Conservation}
\begin{multline}
\label{Flow-Conservation-1}
\sum_{m=1}^{N}\sum_{j=1}^{I}\sum_{i=1}^{I} {X_{m,j,n,i}^{b} L_{b}} + \sum_{o=1}^{O} T_{o,b} L_{b} = \\
    \sum_{o=1}^{O} Y_{o,b} L_{b} + \sum_{i=1}^{I}\sum_{m=1}^{N}\sum_{j=1}^{I} {X_{n,i,m,j}^{b} L_{b}},~~~~~~\forall n, b.
\end{multline}
\begin{multline}
\label{Flow-Conservation-2}
\sum_{m=1}^{N}\sum_{j=1}^{I}\sum_{i=1}^{I} {X_{m,j,n,i}^{b} L_{b}} + \sum_{p=1}^{P} T_{p,b} L_{b} = \sum_{p=1}^{P} Y_{p,b} L_{b} \\
    + \sum_{i=1}^{I}\sum_{m=1}^{N}\sum_{j=1}^{I} {X_{n,i,m,j}^b L_{b}} + \sum_{p=1}^{P} W_{p,b} \theta,~~~\forall n, b.
\end{multline}
\end{subequations}

\textbf{Link Capacity:} Constraint~\eqref{Link-Capacity-1} indicates that the tasks transmission rate does not exceed the link capacity.
\begin{equation}
\label{Link-Capacity-1}
\sum_{b=1}^{B} X_{n,i,m,j}^b\rho_{n,i,m,j}^{b}R_{n,m} \leq \xi R_{n,m},~~~~~~~~~~\forall n, i, m, j,
\end{equation}
where coefficient $\xi$ ensures that the links will not be fully saturated in order to avoid latency due to excessive queuing at the BSs.

\textbf{Server Capacity:} A BS can process the task if and only if it is co-located with a server as it is realized with constraint~\eqref{Server-Capacity-1}. Additionally, constraint~\eqref{Server-Capacity-2} indicates that a server can process the task subject to its capacity.
\begin{equation}
\label{Server-Capacity-1}
Y_{n,b} \leq \Pi_{n},~~~~~~~~~~~~~~~~~~~~~~~~~~~~~~~~~~~~~~~~~~~~~\forall n, b.
\end{equation}
\begin{equation}
\label{Server-Capacity-2}
\sum_{b=1}^{B} Y_{n,b} L_{b} \leq C_{n},~~~~~~~~~~~~~~~~~~~~~~~~~~~~~~~~~~~~~~~\forall n.
\end{equation}

Additionally, to ensure the correct task transmission over the mmWave backhaul, constraint~\eqref{Server-Capacity-3} indicates that a BS serves a given task if the task is received by any interface of that BS \textit{or} the task is initially located at the BS.
\begin{equation}
\label{Server-Capacity-3}
Y_{n,b} \leq \sum_{m=1}^{N}\sum_{j=1}^{I}\sum_{i=1}^{I}\ X_{m,j,n,i}^{b} + T_{n,b},~~~~~~~~~~~~~~~~~\forall n, b.
\end{equation}

\textbf{Cloud Constraints:} As we assume that the cloud has unlimited available resources to serve tasks, the capacity is not a constraint for the cloud. However, to process the task in the cloud, the task has to traverse a path with enough capacity to reach the cloud. The task reaches the cloud if and only if it is previously received by (or initially located at) a BS that is connected to the cloud: 
\begin{equation}
\label{Cloud-Constraints-1}
\sum_{p=1}^{P} W_{p,b} \leq \sum_{m=1}^{N}\sum_{j=1}^{I}\sum_{i=1}^{I} X_{m,j,n,i}^{b},~~~~~~~~~~~~~~~~~~~~~~~~\forall n, b.
\end{equation}

Constraint~\eqref{Cloud-Constraints-2} correlates the decision variables $Y$ and $W$ and indicates that the cloud may process the task only when the cloud receives it. 
\begin{equation}
\label{Cloud-Constraints-2}
Y_{N+1,b} \leq \sum_{p=1}^{P} W_{p,b},~~~~~~~~~~~~~~~~~~~~~~~~~~~~~~~~~~~~~~~~~~~~~~~~\forall b.
\end{equation}

\textbf{Process All Tasks:} Constraint~\eqref{Process-All-Tasks-1} ensures that all tasks must be processed as follows:
\begin{equation}
\label{Process-All-Tasks-1}
\sum_{n=1}^{N} Y_{n,b} = 1,~~~~~~~~~~~~~~~~~~~~~~~~~~~~~~~~~~~~~~~~~~~~~~\forall b.
\end{equation}

\textbf{Initial Constraints:} Initially, mobile users offload their tasks to their corresponding BSs. For BSs that are not connected to the cloud, constraint~\eqref{Initial-Constraint-1} forces the task to either be proceeded at the BS if it is co-located with a server or leave the BS. The BSs that have direct connection to the cloud have this extra possibility to send tasks to the cloud as it is expressed in constraint~\eqref{Initial-Constraint-2}.
\begin{subequations}
\label{Initial-Constraint}
\begin{align}
&  \sum_{n=1}^{N}\sum_{i=1}^{I}\sum_{m=1}^{N}\sum_{j=1}^{I} {X_{n,i,m,j}^b} + \sum_{o=1}^{O} Y_{o,b} = 1,&\forall b. \label{Initial-Constraint-1}\\
&  \sum_{n=1}^{N}\sum_{i=1}^{I}\sum_{m=1}^{N}\sum_{j=1}^{I} {X_{n,i,m,j}^b} + \sum_{p=1}^{P} Y_{p,b} + \sum_{p=1}^{P} W_{p,b} = 1,&\forall b. \label{Initial-Constraint-2}
\end{align}
\end{subequations}

\subsection{Objective Functions}
The ultimate goal is to serve offloaded tasks with minimum latency. The latency that tasks experience considers the following parameters: 
\begin{enumerate}
\item \textbf{Transmission Delay:} The transmission delay for a task in the backhaul network is influenced by the path that the task traverses (defined by $X_{n,i,m,j}^{b}$) in conjunction with the allocated bandwidth to the task ($\rho_{n,i,m,j}^b R_{nm}$).

\item \textbf{Cloud Latency:} A task experiences a notable transmission delay when it is served in the cloud ($W_{p,b} = 1$).
\end{enumerate}
 
Depending on the transmission technique about the backhaul network, the latency can be calculated in two ways. In the first assumption, BSs transmit tasks in a store-and-forward manner. In the sequel, we call this way of task transmission as hop-by-hop (denoted by $hbh$). In this case, the total latency corresponds to the sum of the transmission latencies of each link used by the task\footnote{The queuing latency is another parameter that affects the serving time of a task. However, the queuing latency is not considered in this paper as we are essentially interested in the transmission delay. We will investigate this more elaborate problem in the future extension of this work.}. The $hbh$ case is an ideal scenario and used as a benchmark in this paper. Hence, the latency of a task can be written as follows: 
\begin{equation}
\label{latb-1}
\text{(A):}~\mathcal L_{b}^{hbh} = \sum_{n=1}^{N}\sum_{i=1}^{I}\sum_{m=1}^{N}\sum_{j=1}^{I} \frac {L_b X_{n,i,m,j}^b}{\rho_{n,i,m,j}^b R_{nm}}
	+ \sum_{p=1}^{P} W_{p,b} \theta.
\end{equation}

In the second transmission technique, tasks are forwarded based on the minimum rate of their path similar to decode-and-forward relaying strategy. We call this way of task transmission as minimum rate transmission (denoted by $minR$). We denote by $N^{hops}_{b}$ the total number of links that a task traverses and its expression is given as follows:
\begin{equation}
N^\text{hops}_{b} = \sum_{n=1}^{N}\sum_{i=1}^{I}\sum_{m=1}^{N}\sum_{j=1}^{I} X_{n,i,m,j}^b.
\end{equation}

We also denote by $\Omega_{b}$ the set containing the links traversed by $b$. As a result, in mmWave backhaul network with minimum rate transmission, the latency can be expressed as follows:
\begin{equation}
\label{latb-2}
\text{(B):}~\mathcal L_{b}^{minR} = \frac {L_b}{\frac {1}{N^{hops}_{b}}~\underset{(n,i,m,j) \in \Omega_{b}}{\min}\rho_{n,i,m,j}^b R_{nm}}
	+ \sum_{p=1}^{P} W_{p,b}\theta.
\end{equation} 

Note that calculating the latency in the backhaul network with minimum rate transmission is a more realistic assumption. However, we consider both aforementioned methods of latency calculation in the evaluation section. Finally, the total latency is expressed as follows using~\eqref{latb-1} or~\eqref{latb-2}:
\begin{equation}
\label{objective}
Latency = \sum_{b=1}^{B} \gamma_{b} \mathcal L_{b}^{\mathcal X},
\end{equation}
where $\mathcal X \in \{hbh,minR\}$ and $\gamma_{b}$ is a weight parameter controlled by the operator ($0 \leq \gamma_{b} \leq 1$ and $\sum_{b = 1}^{B} \gamma_{b} = 1$). High value of $\gamma_{b}$ gives more priority to task $b$.

\subsection{Optimization Problem}
The optimization problem aiming at minimizing the total task serving time is formulated as follows:
\begin{align}
\text{(P):}&\quad{\text{minimize}}\quad Latency \label{non-convex-problem}\\
&\quad\text{subject to:}~~\eqref{Interface-Connectivity-1},\eqref{Interface-Connectivity-2},\eqref{Task-Link-1},\eqref{Task-Link-2},\eqref{Flow-Conservation},\eqref{Link-Capacity-1},\eqref{Server-Capacity-1}, \notag\\
&\quad\eqref{Server-Capacity-2},\eqref{Server-Capacity-3},\eqref{Cloud-Constraints-1},\eqref{Cloud-Constraints-2},\eqref{Process-All-Tasks-1},~\text{and}~\eqref{Initial-Constraint}. \notag
\end{align}

The optimization problem (P) is categorized as MINLP due to the fraction in the objective function in addition to the products of decision variables in constraint~\eqref{Link-Capacity-1}.

\section{Proposed Approaches}
\label{sec:approaches}
As it is difficult to optimally solve the optimization problem (P), we propose the following iterative algorithm:

\textit{Step 1: Find Topology:} The linearization of the objective function and constraints enables the formulation of a MILP problem which can be optimally solved. In this step, we assume that the links between mmWave BSs are equally shared between all tasks that traverse the same link. Therefore:
\begin{equation}
\rho_{n,i,m,j}^{b} = \frac {1}{\sum_{b=1}^{B} X_{n,i,m,j}^b},~~~~~~~~~~~~~~~~~~~\forall n,i,m,j.
\end{equation}

For instance, if three tasks are using the link $(n,i,m,j)$ then each task will have a third of the bandwidth. Moreover, to linearize the objective function, we introduce a new decision variable denoted by $Z_{n,i,m,j}$ such that:
\begin{equation}
Z_{n,i,m,j} = \sum_{b=1}^{B} X_{n,i,m,j}^{b},~~~~~~~~~~~~~~~~~~~~~~\forall n,i,m,j.
\end{equation}

By replacing $\rho_{n,i,m,j}^{b}$ with $1/Z$ in (A), the objective function contains the product of a binary ($X_{n,i,m,j}^{b}$) and a continuous variable ($Z_{n,i,m,j}$). We define decision variable $U_{n,i,m,j}^{b}$ as follows:
\begin{equation}
U_{n,i,m,j}^{b} = X_{n,i,m,j}^{b} Z_{n,i,m,j}.
\end{equation}

Finally, the following constraints are added to guarantee the linearity of the problem:
\begin{subequations}
\label{LinearSconst}
\begin{align}
& U_{n,i,m,j}^{b} \leq \bar{Z} X_{n,i,m,j}^{b},~~~~~~~~~~~~~~~~~~~~~~\forall n,i,m,j,\\
& U_{n,i,m,j}^{b} \leq Z_{n,i,m,j},~~~~~~~~~~~~~~~~~~~~~~~~\forall n,i,m,j,\\
& U_{n,i,m,j}^{b} \geq Z_{n,i,m,j}-(1-X_{n,i,m.j}^{b}) \bar{Z},~~\forall n,i,m,j,\\
& U_{n,i,m,j}^{b} \geq 0,~~~~~~~~~~~~~~~~~~~~~~~~~~~~~~~~\forall n,i,m,j,
\end{align}
\end{subequations}
where $\bar{Z}$ is an upper bound of $Z_{n,i,m,j}$. As a result, the objective function (A) is linearized as follows:
\begin{equation}
\mathcal L_{b}^{hbh} = \sum_{n=1}^{N}\sum_{i=1}^{I}\sum_{m=1}^{N}\sum_{j=1}^{I} \frac {U_{n,i,m,j}^b L_b}{R_{nm}} + \sum_{p=1}^{P} W_{p,b} \theta.
\end{equation}

Therefore, the optimization problem can be expressed as follows:
\begin{align}
\text{(P1):}&\quad{\text{minimize}}\quad Latency=\sum_{b=1}^{B} \gamma_{b} \mathcal L_{b}^{hbh} \label{uniform-problem} \\
&\quad\text{subject to}~~\eqref{Interface-Connectivity-1},\eqref{Interface-Connectivity-2},\eqref{Task-Link-1},\eqref{Task-Link-2},\eqref{Flow-Conservation},\eqref{Server-Capacity-1}, \notag \\
&\quad\eqref{Server-Capacity-2},\eqref{Server-Capacity-3},\eqref{Cloud-Constraints-1},\eqref{Cloud-Constraints-2},\eqref{Process-All-Tasks-1},~\text{and}~\eqref{Initial-Constraint} \notag.
\end{align}

Note that the constraint~\eqref{Link-Capacity-1} is removed as, by construction, $Z_{n,i,m,j}$ meets the link capacity constraint. The optimization problem (P1) can be optimally solved using off-the-shelf software. Note that the MILP problem is NP-hard. However, the complexity does not impose a significant concern as we are dealing with a planning approach that does not require real-time solutions.

\begin{figure*}[t]
    \centering\vspace{+2mm}
    \begin{subfigure}[t]{0.475\textwidth}
        \centering
        \includegraphics[width=\textwidth]{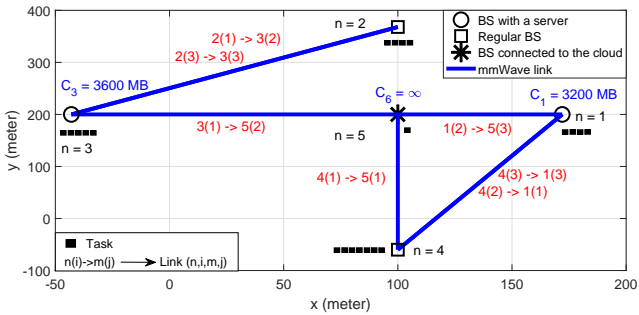}
        \caption[]%
        {{\small N = 5, I = 3, B = 20}}  
        \label{fig:topo-1}
        \vspace{-2mm}
    \end{subfigure}
    \hfill
    \begin{subfigure}[t]{0.475\textwidth}  
        \centering 
        \includegraphics[width=\textwidth]{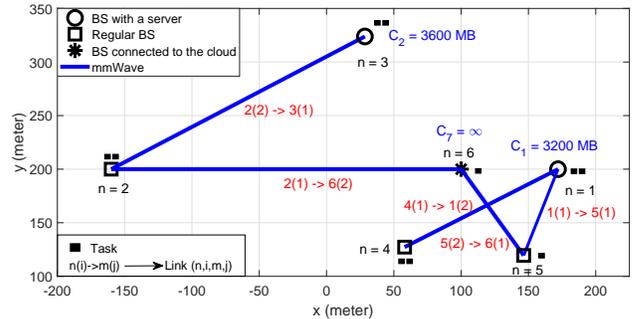}
        \caption[]%
        {{\small N = 6, I = 2, B = 10}}
        \label{fig:topo-2}
        \vspace{-2mm}
    \end{subfigure}
    \caption[]
    {\small Two examples of the optimized MEC infrastructure topologies. The established links $(n,i,m,j)$ are identified as $n(i)\rightarrow m(j)$. The tasks are randomly distributed as indicated with the black filled squares.} 
    \label{fig:topologies}
\end{figure*}

\textit{Step 2: Bandwidth Allocation:} Once the MILP optimization problem (P1) is solved, its output determines the backhaul network topology, the path that each task ought to traverse, and identifies which server is used to process a given offloaded task. Therefore, we can now optimize the  decision variable $\rho_{n,i,m,j}^{b}$ representing the fractions of allocated bandwidth to each task for each backhaul link given the values of $X_{n,i,m,j}$, $X_{n,i,m,j}^{b}$, $Y_{n,b}$, and $W_{n,b}$. For backhaul network with $hbh$ transmission, the following quasi-convex problem finds the optimal bandwidth allocation for each task: 
\begin{align}
\text{(P2A):}&\quad{\text{minimize}}\quad Latency \label{non-problem-1} \\
&\quad\text{subject to}~~\eqref{Link-Capacity-1}. \notag
\end{align}

In (P2A), $\mathcal L_{b}^{hbh}$ is calculated according to~\eqref{latb-1} and its output optimizes the bandwidth allocation for a task on \textit{each link} separately. 

For backhaul network with $minR$ transmission, we define the continuous variable $\Psi_{b}$ as follows:
\begin{equation}
\Psi_{b} = ~\underset{(n,i,m,j) \in \Omega_{b}}{\min}(\rho_{n,i,m,j}^b R_{nm}).
\end{equation}

Then, (B) can be written as follows:
\begin{equation}
\text{(B):}~\mathcal L_{b}^{minR} = \frac {L_b}{\frac {1}{N^{hops}_{b}} \Psi_{b}} + \sum_{p=1}^{P} W_{p,b}\theta.
\end{equation}

Finally, the bandwidth allocation problem for $minR$ transmission is converted into a quasi-convex one as follows: 
\begin{align}
\text{(P2B):}&\quad{\text{minimize}}\quad Latency \label{non-problem-2} \\
&\quad\text{subject to}~~\eqref{Link-Capacity-1}, ~\Psi_{b} \leq \rho_{n,i,m,j}^{b} R_{n,m},\notag
\end{align}
where $\mathcal L_{b}^{minR}$ is calculated according to~\eqref{latb-2}. Once (P2B) is solved, its output determines the optimal bandwidth allocation for a task over a \textit{path} that a task traverse in the mmWave backhaul network until it is processed. 

It is easy to deduce that the problems (P2A) and (P2B) are quasi-convex with respect to $\rho_{n,i,m,j}^{b}$ since their objective functions are the sum of hyperbolas and $\rho_{n,i,m,j}^{b} > 0$ (and $\Psi_b > 0$) and the constraints are linear. These problems can be efficiently solved using the bi-section methods or adaptive subgradient techniques~\cite{doi:10.1080/02331934.2016.1189551, article}.


\section{Numerical Results}
\label{sec:eval}
\begin{figure*}[t]
    \centering
    \begin{subfigure}[b]{0.475\textwidth}
        \centering
        \includegraphics[width=\textwidth]{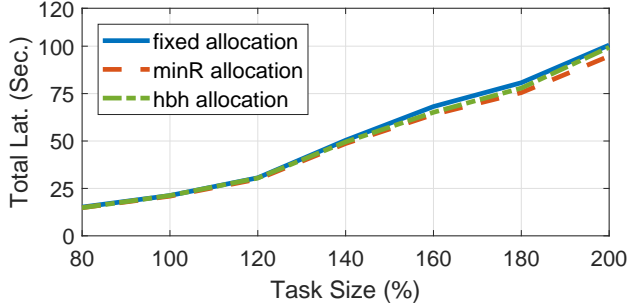}
        \caption[]%
        {{\small hbh transmission}}  
        \label{fig:result-IQ}
        \vspace{-2mm}
    \end{subfigure}
    \hfill
	\begin{subfigure}[b]{0.475\textwidth}   
        \centering 
        \includegraphics[width=\textwidth]{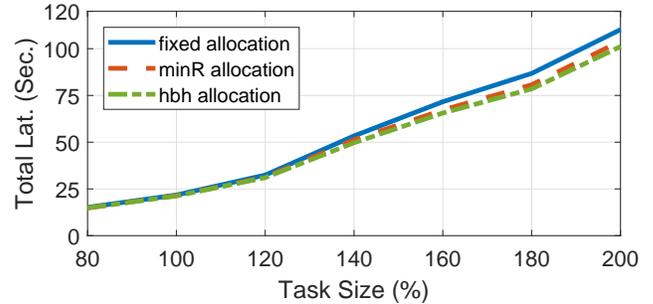}
        \caption[]%
        {{\small minR transmission}}    
        \label{fig:result-FQ}
        \vspace{-2mm}
    \end{subfigure}
    \caption[]
    {\small Total serving time versus the volume of user tasks. Comparison between different bandwidth allocation policies applied for the \textit{hbh} and \textit{minR} metrics.} 
    \label{fig:total-latency}
\end{figure*}

We perform a numerical evaluation to study the behavior of our proposed approach for various system parameters. The MILP problem is solved using the MATLAB toolbox YALMIP~\cite{1393890} employed with the mathematical programming solver GUROBI~\cite{gurobi}. 

BSs are distributed within a 280 $\times$ 280 $\text{m}^{2}$. Tasks are bandwidth-intensive with different sizes between 0.1 to 1 GBytes and they are initially distributed randomly among BSs. In all simulations, only one BS, BS $N$, has a direct connection with the central cloud. The corresponding latency is set to 200 ms~\cite{7414063}. We also assume that two BSs, BS $1$ and BS $3$, are co-located with edge servers having 3.2 and 3.6 GBytes of storage capacity, respectively.

%
%

\subsection{Backhaul Topology and Task Routing Paths}
In this simulation, we visualize the output of our optimization model for two different scenarios. The first scenario considers 5 BSs equipped with 3 interfaces. Twenty tasks will be assigned to the existing MEC infrastructure. The second scenario has 6 BSs with 2 interfaces each and 10 tasks to serve. Fig.~\ref{fig:topologies} depicts the obtained topologies after the execution of our proposed approach. We display the BSs which are co-located with servers using circles, regular BSs with squares, and the BS that has wired connection to the central cloud with the asterisk. The number of black filled squares near each node indicates the number of tasks that have been initially located in that BS. 

In Fig.~\ref{fig:topo-1}, the obtained topology is designed such that the total serving time is minimized. First, the central node connected to the cloud uses its three interfaces to establish connections with three different nodes (i.e, BSs $1$, $3$, and $4$). Indeed, due to the high number of tasks and the limited capacity of the edge servers, few tasks are processed in the MEC while the rest must be sent to the cloud. We can also notice that some BSs establish two concurrent connections with each other (e.g., BSs $1$ and $4$) as there are redundant interfaces that can be exploited to avoid the use of the same mmWave links. Similar remarks are noticed for the second scenario given in Fig.~\ref{fig:topo-2}. However, due to the limited number of interfaces, the topology and the routing paths are optimized such that the overall serving time is minimized. This explains the fact that a link is not directly established between BSs $4$ and $6$. Instead, BS $4$ prefers to establish a link with BS $1$ and then if it has extra tasks, they will be forwarded through BS $5$. The central node cannot establish connections with more than two nodes. Therefore, it is important to intelligently select the nodes to be connected to so that the objectives of the network are met. 

Table~\ref{tab:task-path-latency} shows the serving details of the tasks for the second scenario (in Fig.~\ref{fig:topo-2}). It is shown that three tasks are served in the same BS (tasks $2$, $3$, and $4$) so they experience zero latency (the traversed paths for such tasks are marked with asterisks), while four tasks are sent to the cloud (tasks $5$, $7$, $9$, and $10$). The remaining tasks are offloaded to local edge servers. The corresponding latencies reflect the quality of mmWave links in addition to the size of the tasks. It is also worth mentioning that task $7$ is offloaded to the cloud in spite of being initially located in a BS co-located with a server (BS $1$). This is due to the fact that the size of this task is relatively high. Meanwhile, BS $1$ serves two external tasks coming from BS $4$ to contribute in reducing the total latency.

\begin{table}[t]
\centering
\caption{\label{tab:task-path-latency}Tasks serving details of the second topology.}
\begin{tabular}{C{0.75cm}|C{0.75cm}|C{1.5cm}|C{2.0cm}|C{1.0cm}}
\hline
Task \# & Size (GB) & Initial Location & Path & Latency (Sec.) \\ \hline
1 & 1.28 & 2 & 2(2)$\rightarrow$ 3(1) & 1.27 \\ \hline
2 & 1.60 & 1 & * & 0 \\ \hline
3 & 0.58 & 3 & * & 0 \\ \hline
4 & 1.37 & 3 & * & 0 \\ \hline
5 & 1.85 & 5 & 5(2)$\rightarrow$ 6(1) 6(3)$\rightarrow$ Cloud & 1.34 \\ \hline
6 & 0.45 & 4 & 4(1)$\rightarrow$ 1(2) & 0.43 \\ \hline
7 & 1.96 & 1 & 1(1)$\rightarrow$ 5(1) 5(2)$\rightarrow$ 6(1) 6(3)$\rightarrow$ Cloud & 1.86 \\ \hline
8 & 0.91 & 4 & 4(1)$\rightarrow$ 1(2) & 0.61 \\ \hline
9 & 0.90 & 6 & 6(3)$\rightarrow$ Cloud & 0.20 \\ \hline
10 & 1.27 & 2 & 2(1)$\rightarrow$ 6(2) 6(3)$\rightarrow$ Cloud & 1.95 \\ \hline
\end{tabular}
\vspace{-2mm}
\end{table}

\subsection{Total Serving Time}
This simulation investigates the performance of our proposed approach versus different task sizes using the scenario given in Fig.~\ref{fig:topo-1}. The results are depicted in Fig.~\ref{fig:total-latency}. Herein, as expected, \textit{hbh} and \textit{minR} achieve the best total latency when applied to their respective transmission schemes. However, it is important to notice that the gap between both approaches remains very small regardless of the objective function. Hence, we conclude that the \textit{minR} transmission-based allocation presents acceptable performances close to those of the benchmark solution. On the other hand, its gap with the fixed bandwidth allocation policy increases especially for a high volume of the tasks. Indeed, starting from a task size of 120\%, the bandwidth allocation solution
outperforms the fixed one to reach 10\% for high task sizes.

On the other hand, we notice that, as expected, the total latency increases with the increase of the size of the user tasks. However, the trend of this increase becomes more important starting from 100\%. This is due to the fact that, for task size less than this value, most of the tasks are processed in edge servers. However, for higher values, the tasks are offloaded to the cloud server since the capacity of the edge servers cannot host many tasks.

\subsection{MEC and Path Diversity}
\begin{figure}[t]\vspace{+2mm}
\begin{centering}
\includegraphics[width=8.5cm]{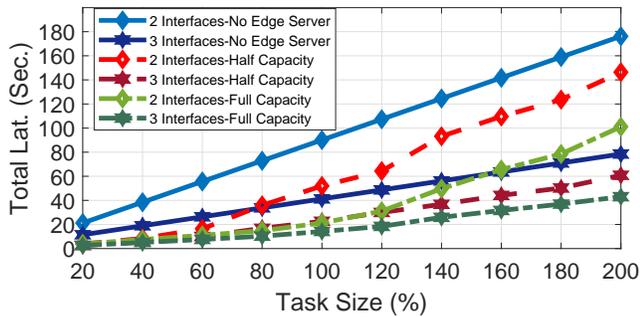}
\par\end{centering}
\caption{\label{result-2} Total latency versus task size for different number of interfaces and edge server capacities.}
\vspace{-2mm}
\end{figure}

This simulation investigates the effect of MEC infrastructure and the number of interfaces in the mesh backhaul network on the total task serving time using the scenario given in Fig.~\ref{fig:topo-1}. We increase the number of BS interfaces from 2 to 3 and consider three different cases for the server capacity (zero, half, and full capacity). In this simulation, the optimization problem with minimum rate (\textit{minR}) transmission is used to calculate the total serving time. As it is shown in Fig.~\ref{result-2}, when BSs are equipped with more interfaces the total latency is significantly decreased. For instance, it is decreased by more than 50\% for the half capacity scenario and 100\% of the task size. The main reason is that increasing the number of interfaces provides more flexibility to the optimizer to find other routing paths and reduces the share of bandwidth between user tasks. Likewise, when servers at MEC infrastructure can serve more tasks then, the total serving time of all tasks is significantly decreased. Indeed, for the zero capacity scenario, all tasks go to the central cloud. 

\section{Conclusion and Future Work}
\label{sec:conclusion}
This paper investigated the problem of task allocation in mmWave mesh backhaul networks. A generic MINLP optimization problem is developed aiming at minimizing the total task serving time. A two-step approach, involving a MILP and a quasi-convex problem is developed to solve the optimization problem. First, the output of the MILP problem determines an optimal network topology, identifies which server is used to process a given offloaded task, and finds the routing paths of each user task. Later, the quasi-convex problem further optimizes the fractions of allocated bandwidth on each mmWave backhaul link. We compared the values of task serving time after each step and investigated the effect of MEC infrastructure and the number of BS interfaces on the system performance. A numerical evaluation showed that the total task serving time is significantly decreased when the bandwidth allocation is optimized, particularly for bandwidth-intensive tasks. Additionally, the evaluation results illustrated a notable decrease of the latency when MEC infrastructure exists at the backhaul networks especially when BSs are equipped with more interfaces. In the future extension of this work, we intend to propose low complex heuristic approaches to solve this problem and incorporate the latency due to queuing.



\section*{Acknowledgment}

This work has been supported by the Knowledge Foundation Sweden through the SOCRA project.

{\small{}\bibliographystyle{./IEEEtran}
\bibliography{refs}
}{\small \par}

\end{document}